\documentclass[aps,prl,twocolumn,showpacs,groupaddress,
superscriptaddress
]{revtex4-1}
\usepackage{bm}
\usepackage{amsmath}
\usepackage{amssymb}
\usepackage{graphicx}

\def\br{{\boldmath r}}

\input{tcilatex}
\pdfoutput=1
\begin{document}

\title{Coulomb interaction and first order superconductor-insulator
transition
}
\author{S.V.Syzranov}
\affiliation{
Theoretische Physik III, Ruhr-Universit\"{a}t Bochum, D-44801 Bochum,
Germany}
\author{I.L. Aleiner}
\author{B.L. Altshuler}
\affiliation{Physics Department, Columbia University, New York,
  N.Y. 10027, USA}
\author{K.B. Efetov}
\affiliation{
Theoretische Physik III, Ruhr-Universit\"{a}t Bochum, D-44801 Bochum,
Germany}\affiliation{Physics Department, Columbia University, New York,
  N.Y. 10027, USA}
\date{\today }

\begin{abstract}

The superconductor-insulator transition (SIT) in  regular arrays of Josephson
junctions is studied  at low temperatures. Near the
transition a Ginzburg-Landau type action containing the imaginary time is
derived. The new feature of this action is that it contains a gauge field $%
\Phi $ describing the Coulomb interaction and changing the standard critical
behavior. The solution of renormalization group (RG) equations derived at
zero temperature $T=0$ in the space dimensionality $d=3$ shows that the SIT
is always of the first order. At finite temperatures, a tricritical point
separates the lines of the first and second order phase transitions.
The same conclusion holds for $d=2$ if the mutual capacitance is larger
than the distance between junctions.
\end{abstract}


\maketitle
{\em Introduction} --After several decades of intensive studies
Josephson junctions arrays (JJA) still remain an important
and inspiring problem. These, at first glance,  simple systems 
exhibit, depending on parameters, superconducting, insulating or metallic
properties (see for a review \cite{fazio}). As a model, JJA
is relevant for granular superconductors and disordered
superconducting films \cite{gantmakher}.

Coulomb interaction (CI) is crucial for
properties of the JJA. It suppresses the density
fluctuations, {\em i.e.} causes  fluctuations of phase $\varphi$ of the
superconducting order parameter, which can destroy the superconductivity even at
zero temperature, $T=0$. 
As the charging energy of one grain increases, the JJA undergoes a
superconductor-insulator transition (SIT) -- an example of
a quantum phase transition. As we show in this Letter the long-range
nature of the CI qualitatively 
affects the SIT.

\begin{subequations}
\label{model}JJA can be described by the effective Hamiltonian \cite{efetov1980} 
\begin{equation}
\hat{H}=\frac{1}{2}\sum_{\br,\br^{\prime }}B_{|\br-\br^{\prime }|}\hat{n}_{\br}\hat{n}%
_{\br^{\prime }}-\sum_{\br,\br^{\prime }}J_{\br,\br^{\prime }}\cos \left( \varphi
_{\br}-\varphi _{\br^{\prime }}\right) ,\quad  \label{a1}
\end{equation}%
where $J_{\br,\br^{\prime }}$ is
the Josephson energy of the junction between neighboring
grains $\br$ and $\br^{\prime },$ 
and $\hat{n}_{\br}$ is the particle number operator
$\hat{n}_{r}=-i\partial /\partial \varphi _{\br}$ in a grain. We
assume that superconducting pairing within each grain is strong and
 consider the limit of the infinite  single-electron gap. Accordingly, the
excitations of isolated grains have the charge $+2e/(-2e)$ (bosons/antibosons). 
Matrix
$B_{r-r^{\prime }}$ describes the interaction between two bosons
on the metallic grains  $r,r^{\prime }$. It is well approximated as
\begin{equation}
B_{r}^{d=3}\simeq \frac{e^2\ell}{\pi{\cal C}r};
\quad 
B_{r}^{d=2}\simeq
\frac{2e^2}{\pi{\cal C}}\ln
\left(\frac{2\pi{\cal C}+r}{r}\right).
\label{B}
\end{equation}
Here $\ell$ is the period of JJA, and ${\cal C} \gtrsim \ell $ is the mutual
capacitance of the neighboring grains.
The energy to charge one grain can be estimated as $B_0\simeq
B_{r\sim \ell}$.
\end{subequations}

Equation (\ref{a1}) looks like a Hamiltonian of a quantum $XY$
-model whose critical behavior is described by an $N=2$
-component $\phi ^{4}$ field theory. Such a field theory was
discussed, e.g., in Refs.~\cite{cha,otterlo,sachdev}  for both finite $T$
and $T=0$. For $T>0$, the critical behavior near the phase
transition is described by a $d$-dimensional $2$-component $\phi
^{4}$ theory:  the quantum transition at $T=0$
involves imaginary time $\tau $ as an additional dimension, {\em i.e.} the same $\phi ^{4}$
 theory should be considered in $d+1$ dimensions. In both cases the SIT is of the second order with the 
critical behavior of $XY$ model in $d$ or $d+1$ dimensions.

This is  correct  if the matrices $B_{\br-\br^\prime}$ in Eq. (\ref%
{B}) are either diagonal, $B_{\br,\br^{\prime }}^{\left( 0\right) }=B_{0}\delta
_{\br,\br^{\prime }}$ or sufficiently short ranged. The effect of the
long range part of Eq.~(\ref{B}) on the critical
behavior has not been investigated so far.

In this Letter we derive a field theory that properly describes the
SIT at low temperatures in the disorder-free model 
(\ref{model}). The Coulomb interaction ($1/\left\vert \br-\br^{\prime }\right\vert $
decay of $B_{r,r^{\prime }}$ at large distances) results in an additional
gauge field $\Phi $ in the Ginzburg-Landau (GL) expansion near SIT and, 
at $T\to 0 $, causes additional logarithmic divergences in the upper critical dimensionality $d=3$ of
JJA. We derive and solve renormalization group (RG)
equations. Solutions demonstrate the first order SIT
at sufficiently low temperatures. To describe SIT in $2D$ JJA we
 use $\epsilon $-expansion at small $\epsilon =3-d$. We find first order SIT
as long ${\cal C} \gg \ell$, and it may become continuous otherwise.

{\em Problems with mean field description}--
To derive the field theory of fluctuations near the
phase transitions one should write a GL expansion in 
superconducting order parameter $\Delta =J\left\langle \exp \left( i\varphi
_{r}\right) \right\rangle ,$ where $\left\langle ...\right\rangle $ is the
quantum mechanical average with the Hamiltonian  (\ref{a1}),
and $J=\sum_{r}J_{r\mathbf{,}r^{\prime }}$. The mean field approximation is
obtained by minimizing this expansion. At first glance, the
free energy functional $F\left[ \Delta \right] $ can be derived straightforwardly and 
should have a form of the standard GL expansion.
Indeed, $F\left[ \Delta \right] $ for time-independent $%
\Delta _{r}$ has the form%
\begin{subequations}
\begin{equation}
F\left[ \Delta \right]
=\sum_{\br,\br^\prime}\left[\alpha_{\br,\br^\prime}
\Delta_{\br}\Delta_{\br^\prime}^*
+\beta_{\br,\br^{\prime }}\left\vert \Delta _{\br}\right\vert
^{2}\left\vert \Delta _{\br^{\prime }}\right\vert ^{2}\right],  \label{b1}
\end{equation}%
where the phase transition is controlled by
\begin{equation}
\alpha_{\br,\br^\prime}=\left[J^{-1}\right]_{\br,\br^\prime}
-{2}\delta_{\br,\br^\prime}/E_0,\quad  \label{a15}
\end{equation}%
and $E_{0}=B_{0}/2$ is the energy of adding or subtracting one Cooper pair
(boson) to a grain.
Apparently, according Eqs.~(\ref{b1})-(\ref{a15}), there exists the critical
coupling $J_{c}$%
\begin{equation}
J_{c}=E_{0}/2  \label{b2},
\end{equation}%
so
that at $J>J_{c}$ the system is a superconductor, while the
insulating state corresponds to $J<J_{c}$. 
However, this conclusion relies on a  quartic term being local and positive. Explicit calculation of
the function $\beta_{\br,\br^{\prime }}$ starting from Eq. (\ref{a1}) gives \cite{eckern}%
\begin{equation}
\beta_{\br,\br^{\prime }}=\frac{1}{E_{0}^{3}}\left[ \frac{7}{2}\delta
_{\br,\br^{\prime }}-4\left( \frac{E_0}{E_{\br,\br^{\prime }}^{-}}+\frac{E_0}{%
E_{\br,\br^{\prime }}^{+}}-1\right) \right],  \label{b3}
\end{equation}%
where $E_{\br,\br^{\prime }}^{\pm }=B_{0}\pm B_{\br,\br^{\prime }}$ are the energies
of two bosons (boson-antiboson) pairs located on the grains $r$ and $%
r^{\prime }$. At large distances $%
B_{r,r^{\prime }}\ll B_{0}$, and Eq.~(\ref{b3}) yields
\begin{equation}
\beta_{|r-r^{\prime }|\gg \ell}=-B_{r,r^{\prime
}}^{2}/E_{0}^{5},  \label{b5}
\end{equation}%
As at large distances $B_{r,r^{\prime }}\propto
\left\vert r-r^{\prime }\right\vert ^{-1}$, the sum over $r^{\prime }$
in Eq.~(\ref{b1})
is negative and diverges linearly (logarithmically) for
coordinate-independent $\left\vert \Delta \right\vert $ in
$3(2)$-dimensional JJAs. 
Such divergences at large distances
signal that the additional soft modes should be included to
make the theory local, and the conventional naive mean-field
is not conclusive.
 This problem arises at low
temperatures, $T\ll E_{0},$ only, while for $T\gg E_{0}$, the Coulomb
interaction can be neglected and 
the SIT temperature in the mean-field approximation \cite{efetov1980}
turns out to be
\end{subequations}
\begin{equation}
T_{c}=J.  \label{b5a}
\end{equation}%
In the vicinity of $T_c$ the conventional GL free energy with a
time-independent $\Delta$ is valid.

{\em Effective field theory}--
To account for the long range interaction we
modify the model slightly: we separate  $B_{r,r^{\prime }}$ 
Eq.~(\ref{B}) into the local  and long-range parts and smoothen the latter:
\begin{equation}
B_{\br}\simeq 2E_0\delta _{\br,0}+{B}_{|\br|+\gamma l},\ 
\label{a2}
\end{equation}
where $\gamma \gtrsim 1$ which controls the short distance cut-off 
that will drop out of  final results.
Although we assume $E_0 \gg \ell\partial_r{B}_{|\br|}|_{r\simeq \ell}$, the
number of long range terms is infinite and their effect accumulated from large
distances has to be included together with quantum fluctuations of $\Delta$.
We separate the Hamiltonian into the bare one, $H_0$, and perturbations $H_{J,B}$
\begin{equation}
\begin{split}
& \hat{H}=\hat{H}_{0}+\hat{H}_{J}+\hat{H}_{B}, \quad 
\hat{H}_{0}=\sum_{\br}E_0\hat{n}_{r}^{2},\\
& \hat{H}_{B}=
\sum_{r,r^{\prime }}\frac{\tilde{B}_{\br,\br^{\prime }}\hat{n}_{\br}\hat{n}%
_{\br^{\prime }}}{2}, \quad 
\hat{H}_{J}=-\sum_{r,r^{\prime }}J_{r,r^{\prime }}\cos \left( \hat{\varphi}
_{r}-\hat{\varphi} _{r^{\prime }}\right),
\end{split}
\raisetag{50pt}
\label{a3}
\end{equation}
and write the partition function
\begin{equation}
{Z}=Tr\ e^{-\hat{H}/T}Z_0\left\langle
T_{\tau }
e^{ -\int_{0}^{1/T}
 d\tau \left[ \hat{H}_{J}\left( \tau\right)
+\hat{H}_{B}\left( \tau \right) 
\right]}
\right\rangle_0.
\label{a8}
\end{equation}
Here  $T_\tau$ stands for imaginary
time ordering, $\left\langle \dots\right\rangle_0
\equiv{Z_0}^{-1}Tr\left\{e^{-\hat{H}_0}/T\dots\right\}$, $Z_{0}=Tr\left\{ e^{ -
  \hat{H}_{0}}/T\right\}$, and $\hat{H}_{B,\, J}\left( \tau \right)
\equiv e^{\hat{H}_{0}\tau } 
\hat{ H}_{B,\, J}e^{ -\hat{H}_{0}\tau}$. 
Terms $\hat{H}_{B}\left( \tau \right) $ and $\hat{H}%
_{J}\left( \tau \right) $ are decoupled by Hubbard-Stratonovich fields $\Delta _{\br}\left(
\tau \right) $ (complex) and $\Phi _{r}\left( \tau \right) $ (real):
\begin{equation}
\begin{split}
&\frac{Z}{Z_0}=\!
\int\!\!{\cal D}\Delta_\br
{\cal D}\Delta ^{\ast
}_\br{\cal D}\Phi_\br e^{-S_{1}}
\left\langle T_\tau e^{-\int_0^{1/T} \hat{H}_{1}d\tau}  \right\rangle_0,\\
& \hat{H}_{1}=\sum_{\br}\Big\{
i\left[\hat{n}_{\br} \Phi _{\br}\right](\tau)
 +\left[\Delta _{\br} e^{-i\hat{\varphi} _{\br}}\right]\left( \tau
\right) 
+\left[\Delta _{\br}^{\ast } e^{i\hat{\varphi}_{\br}}\right]\left( \tau
\right) \Big\},
\\
&S_1=\sum_{r,r^{\prime }}\int_{0}^{1/T } d\tau\left\{ J_{r,r^{\prime
}}^{-1}\left[\Delta _{r}\Delta _{r^{\prime }}^{\ast }\right](\tau)+\frac{1}{2}\tilde{B}%
_{r,r^{\prime }}^{-1}\left[\Phi _{r}\Phi _{r^{\prime }}\right](\tau)\right\},
\end{split}
\raisetag{87pt}
\label{b8}
\end{equation}
where $\sum_{r^\prime}\tilde{B}
_{r,r^{\prime }}^{-1}B_{|r^\prime|+\gamma \ell}=\delta_{\br,0}$.

In the limit of low temperatures $T\ll E_0$  the average $%
\left\langle ...\right\rangle _{0}$ can be calculated over the ground state when all grains
are neutral. We neglect exponentially small contributions like
$\exp \left( - E_{i}/T\right),$ where $E_{i}$ are the eigenenergies of Hamiltonian $\hat{H}_{0}$ 
corresponding to the charged isolated grains. At the same time, we keep finite $1/T$ when
integrating over $\tau $ in Eqs. (\ref{b8}) to obtain
algebraic in $T$ contributions.
This calculation is carried
out near the SIT by cumulant expansion assuming that the fields $\Delta
\left( \tau ,\br\right), \Phi(\tau,\br)$ are slow in both time $\tau $ and coordinate $r$.
As we compute averages with the bare Hamiltonian $\hat{H}_0$, all
generated terms remain local. 
Introducing continuous coordinate description we obtain
\begin{subequations}
\label{a11all}
\begin{equation}
Z \!=\!\int e^{ -\int_0^{1/T} d\tau(L_{\Delta }+L_{\Phi })} {\cal D}\Delta(\br,\tau)
{\cal D}\Delta ^{\ast
}(\br,\tau){\cal D}\Phi(\br,\tau).  
\label{a11} 
\end{equation}
The  order parameter is controlled by Lagrangian
\begin{equation}
\begin{split}
L_{\Delta }=
\int d^dr &\Big\{ 
f^2\Delta^*\left[
a E^2
- \left( \partial_\tau +i\Phi \right)^2-  c^{2}\nabla^2 
\right]\Delta
\\
& + b f^4\pi^2c^d E^{\epsilon}\left\vert \Delta \right\vert ^{4}%
\Big\},
\end{split} \raisetag{20pt}
\label{a11b}
\end{equation}
where $\epsilon\equiv 3-d$.
The coefficients in this action are expressed in terms of the initial constants
of the Hamiltonian (\ref{a3}) at $|\delta J| \ll J_c=E_0/2;
 \delta J=J-J_c$, as%
\begin{equation}
c^{2}\!=\frac{E_0^2\ell^2}{2d}; \ 
a=-\frac{2\delta J}{E_0};\ 
f^{2}\!=\frac{2}{ E_{0}^{3}\ell^d};\
b=\frac{7\!\left({2d}\right)^{d/2}}{8\pi^2}.
  \label{a11a}
\end{equation}
Energy $E$ is the running high-frequency cut-off in the theory, it
starts at $E=E_0$. We included this cut-off explicitly
in the action to keep the interaction constants dimensionless, avoid
rescaling of
$(r,\tau)$  during RG, and explicitly illuminate the
dimensionality of the interaction terms.
As $f$ can be always removed by
the rescaling of $\Delta$, the physical results may depend only on constants $a,b,c$.

The fluctuating voltage $\Phi /\left( 2e\right)$ is
of great importance for the critical behavior.
This field is controlled by the Gaussian Lagrangian
\begin{equation}
L_{\Phi } =\frac{1}{2}\int drdr^{\prime }\Phi\left( \tau,\br
\right)  \left[\tilde{B}^{-1}\right]_{\br,\br^{\prime }}\Phi\left( \tau,\br^\prime \right);
\label{a11c}
\end{equation}
After the coordinate Fourier
transform of the fields $\Phi(k,\tau)=\Phi^*({-k,\tau})$, we obtain
from Eq.~(\ref{B}): 
\begin{equation}L_{\Phi } =\!\int\! \frac{d^dk}{(2\pi)^d}
|\Phi(\tau,k)|^2\left(\frac{g_1|k|^{d-1}}{8\pi^2
    c}+\frac{g_2k^2E_0^{-\epsilon}}{8\pi^2 c^{d-2}}\right).
\label{a11d}
\end{equation}
Harmonics $\Phi(k)$ with  $|k| \gtrsim 1/(\gamma\ell)$ are suppressed.
The  coupling constants in  Eq.~(\ref{a11d}) are
defined as 
\begin{equation}
\begin{split}
&g_1^{d=3}=\left(\frac{\pi^2{\cal C}E_0}{\sqrt{6} e^2}\right);\quad
 g_1^{d=2}=
\left(\frac{\pi \ell E_0}{4 e^2}\right);\\
&g_2^{d=3}=0;\quad g_2^{d=2}=\left(\frac{\pi^2E_0 {\cal C}}{e^2}\right).
\end{split} 
 \label{a11e}
\end{equation}
\end{subequations}
For this approach to be applicable, the fluctuations of $\Phi$ should
be small {\em i.e.} $g_1+g_2 \gtrsim 1$.

The field theory (\ref{a11all}) defined for two slow
fields $\Delta,\Phi$ helps one to avoid  the negative
and non-local quartic terms  Eq. (\ref{b5}), arising in a
single field formulation (\ref{b1}).
Indeed, Eqs.~(\ref{a11all}) are invariant
under the transformation
\begin{equation}
\Phi _{r}\left( \tau \right) \rightarrow \Phi _{r}\left( \tau \right)
+\phi(\tau);
\quad \Delta \to \Delta e^{-i\int_0^\tau d\tau_1 \phi(\tau_1)}
 \label{a20}
\end{equation}
for $\int_0^\beta \phi(\tau) d\tau=0\, (\!\!\!\!\mod 2\pi)$, {\em i.e.} the
effect of the fluctuations of $\Phi_ke^{i\omega\tau}$ at $kc \ll
\omega$ vanishes. Fixing by hand $\Delta$ while allowing for all the fluctuations of $\Phi$ violates the
gauge invariance (\ref{a20}), and overestimates the 
contributions of small $k$ and leads to an incorrect non-local theory. 

{\em Renormalization group analysis}--
To the best of our knowledge, the model (\ref{a11all}) has never been
discussed. We analyze it using the RG
approach in $d=3$ and $d=3-\epsilon$.

The cut-off dependence, $E^\epsilon_0$. in Eqs.~(\ref{a11b}),
(\ref{a11d}) suggests that the theory is logarithmic for $d=3$
and $a=0$. As long as $E_0\gtrsim T$, $\tau$ can be considered as
an extra dimension and the gauge invariance (\ref{a20}) prohibits generating
a relevant term $\propto \Phi^2$. The other possible terms allowed
by symmetry
are irrelevant, {\em i.e.} the theory is re-normalizable.
For $\epsilon=1$, Eqs.~(\ref{a11a}) and (\ref{a11d}) still contain all
the relevant terms. Moreover, term $\propto g_1/c$ describes the long
range Coulomb interaction and cannot be re-normalized. The term
$\propto g_2$ is leading irrelevant and it describes the logarithmic interaction and its renormalization
due to the virtual boson-antiboson pairs.

We subdivide the fields $\Delta $ and $\Phi 
$ into slow $\tilde{\Delta},\tilde{\Phi}$ and fast $\Delta _{0},\Phi _{0}$
parts (In the first loop approximation cut-off procedure can be
rather arbitrary, we treat energy $E_0$ as a running cut-off), integrate in Eq.~(\ref{a11}) over $\Delta _{0},\Phi
_{0}$ making a cumulant expansion in $\tilde{\Delta},\tilde{\Phi}$ up to the fourth order.
The  action is
reproduced with the couplings running as
\begin{subequations}
\begin{eqnarray}
&&\frac{d{a}}{dl} =\left\{a\left(
    2-\frac{{b}}{2}\right)\right\}+\frac{3a}{2}
\frac{1}{{g}_{1}+{g}_{2}};\\ 
&&\frac{d{b}}{dl} =\left\{\epsilon{b}-\frac{5}{4}{b}^{2}\right\}-
\frac{2}{\left({g}_{1}+{g}_{2}\right) ^{2}}+\frac{{b}}
{{g}_{1}+{g}_{2}};  \\
&&\frac{d{c}}{dl} =\frac{2}{3}\frac{c}{{g}_{1}+{g}_{2}};\\ 
&&\frac{d{g}_{1}}{dl}=\frac{2}{3}\frac{{g}_{1}}{{g}_{1}+{g}_{2}};   \\
&&\frac{d{g}_{2}}{dl} =-\epsilon {g}_{2}+\frac{2}{3}\frac{{g}_{2}}{{g}_{1}+{g}_{2}}+\frac{1}{6}  
\end{eqnarray}%
\label{a25}
\end{subequations}
(the running of $f$ is of no consequence).
Here 
\begin{equation}
l=\ln\left({{\bar{B}}}/{E}_0\right),
\label{l}
\end{equation}
and Eqs.~(\ref{a11a}), (\ref{a11e}) are the initial conditions for
Eqs.~(\ref{a25}). The terms in curly brackets correspond to the
$\beta$-function for $XY$ model in $d+1$ dimension.

RG flow should be stopped at $l\gtrsim l_*$,
\begin{equation}
l_*={\rm max}\left(l_t;l_a\right); \ \
\left|a\left(l_a\right)\right|\simeq 1, \ \ l_T\simeq \ln({\bar{B}}/{T}),
\label{E*}
\end{equation}
where the quantum fluctuations loose importance either due to
the finite temperature ($l_*=l_T$) or to departure from the phase
transition line ($l_*=l_a$). In the former case
the quantum RG analysis (in $d+1$ dimensions) should be supplemented
by the analysis of the $d$-dimensional classical fluctuations.

This analysis deserves some discussion.
Let us obtain the free energy  from
Eqs.~(\ref{a11b}) and (\ref{a11d}) by including only time independent
fields:
\begin{equation}
\begin{split}
F&=
\int d^dr \Big\{ c^2_T\left|\nabla \Delta\right|^2
+
a_T T^2\left|\Delta\right|^2
+ b_T\pi^2c^d_T T^{\epsilon}\left\vert \Delta \right\vert ^{4}%
\Big\}\\
&+  \int d^dr \left|\Delta\right|^2\Phi^2
\\
&+
\!\int\! \frac{d^dk}{(2\pi)^d}
\frac{|k|^{d-1}|\Phi(k)|^2}{{4\pi^2
    c_T}}
\left[
g_{1,T}+g_{2,T}\left(\frac{c_Tk}{T}\right)^\epsilon\right]
,
\end{split}
\label{F} \raisetag{50pt}
\end{equation}
where the subscript $T$  means that the couplings are
calculated at $l=l_T$. To obtain the canonical form one has to
integrate $e^{-F/T}$ over all the configuration of $\Phi$.
Such integration would immediately produce the term $\simeq
-|\Delta|^3$ resulting in the first order phase transition.
Similar  effect of the fluctuating magnetic field was studied long
ago\cite{hlm}.

\begin{figure}
\includegraphics[width=0.9 \columnwidth]{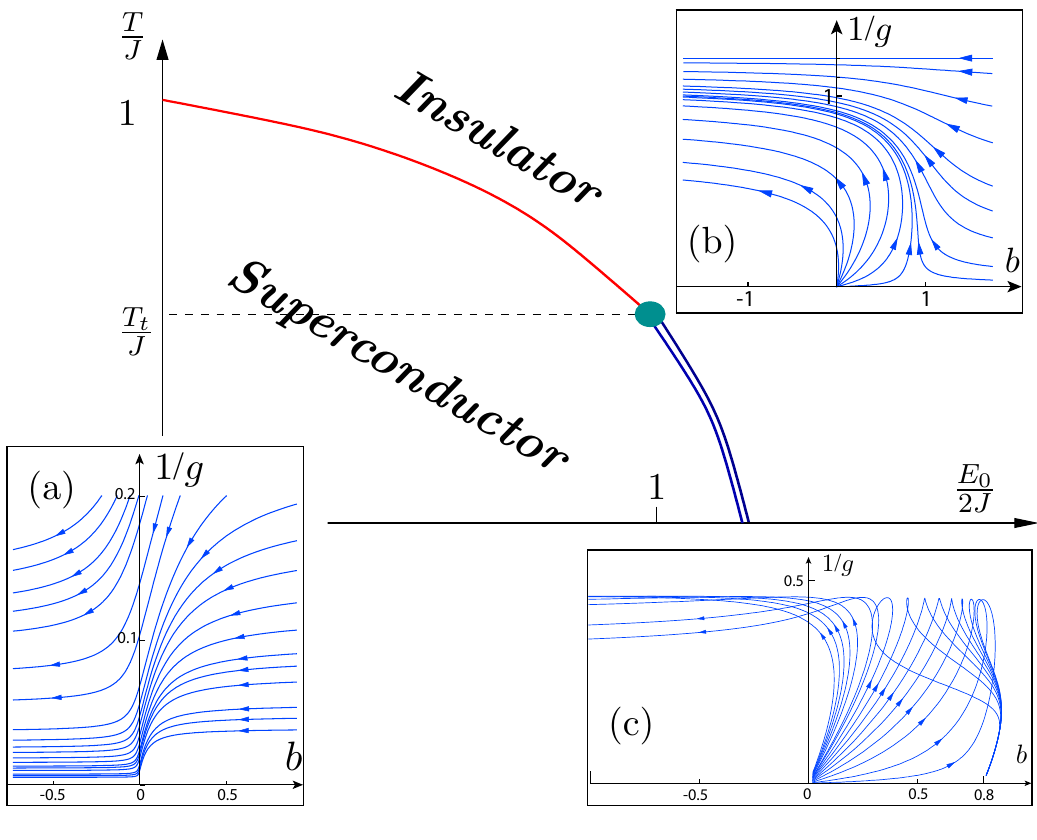}
\caption{Phase diagram of the JJA. The solid and
  double- lines are the second and first order respectively. Insets: (a)
the two parameter RG flow for 3D JJA; (b)  two
parameter RG flow for 2D arrays with logarithmic interaction;
(c) the projection of three parameter RG flow for the logarithmic
and ``moderate'' Coulomb interaction $g_1(l=0)=0.65$.}
\label{fig1}
\end{figure}

However, the electrostatic potential is very different from the
 vector potential of the magnetic field. The gauge invariance
prohibits the screening of the static vector potential by the fluctuations of $\Delta$.
On the contrary, the static electrostatic potential can be screened [gauge invariance
(\ref{a20}) does not allow to remove zero Matsubara component of
$\Phi(\br)$.]
Explicit calculation of the static polarization operator 
adds extra term 
\begin{equation}
F_{scr}=\frac{1}{6T}\left(\frac{T}{c_T}\right)^{d}
\int \left\vert \Phi_{k} \right\vert ^{2} \frac{d^{d}k}{\left( 2\pi \right) ^{d}},
\label{b12}
\end{equation}%
to the free energy $F$ [this term is an effect of finite $\tau$ and
does not appear in RG analysis (\ref{a25})].
As the result, the long-range fluctuations of $\Phi$ are
massive, the last two lines in Eq.~(\ref{F}) can be neglected  and
we are left with a classical $XY$-model in $d$ dimensions. The phase
transition is of the second order in $d=3$ and
Berezinskii-Kosterlitz-Thouless in $d=2$. 
Transition temperature  estimated as
\begin{equation}
T_c= \bar{B} e^{-l_c}; \quad a\left(l_c\right) \simeq
-b\left(l_c\right); 
\label{Tc}
\end{equation}
is shown by solid line on Fig.~\ref{fig1}.

{\em First order phase transition at $d=3$}--
Second order phase transition (\ref{Tc}) implies $b>0$, so that 
the energy 
\begin{equation}
U_T(|\Delta|)\simeq a(l_{\cal E}){\cal E}^2|\Delta|^2+\pi^2c(l_{\cal E})^{d}b(l_{\cal E})|\Delta|^4, \   
\label{U}
\end{equation}
with  ${\cal E}={\rm max}(T,|\Delta|),\ l_{\cal E}=\ln\frac{\bar{B}}{\cal E}$
 has a single minimum.

In what follows, we solve the RG equations (\ref{a25}) and
show that at some value of $l_t$,
$b\left(l_t\right)=0$.
It means, that at $T<T_t=\bar{B}\exp(-l_t)$ the additional stable
minimum appear in Eq.~(\ref{U}); the transition is of the first order -- theory
is massive and  renormalization terminates. The vicinity of $T_t$ can be analyzed in the spirit of Ref.~\cite{wegner}.

For $d=3$ only  ${g}={g}_{1}+{g}_{2}$ is important
and we find
\begin{equation}
\begin{split}
{g}\left( l\right) = \frac{5\xi}{6},
\ \ {b}\left( l\right) =\frac{1}{\xi}{\cal F} \left(
\ln\frac{\xi}{\xi^*} \right),\ \ \xi =l+\frac{6g_{1}^{d=3}}{5},  
\end{split}
\label{b10}
\end{equation}%
where $g_1^{d=3},b$ are given by Eqs.~(\ref{a11a},\ref{a11e}),
 and $\xi^*$ is found from 
$
6g_1^{d=3}b/5={\cal F}[\ln(6g_1^{d=3}/5\xi^*)].
$
 Function 
\begin{equation}
{\cal F}(x)=\frac{22}{25}+\frac{2\sqrt{239}}{25}\cot\left(\frac{\sqrt{239}}{10} x\right)
\label{F}
\end{equation}
changes sign at $x\approx -0.62\dots$.
Therefore, $b$  changes the sign at finite
$l_t$ independently on the initial conditions, see Fig.~\ref{fig1}a. 
As the initial couplings, $g_1^{d=3}\gtrsim 1,\ b\lesssim 1$
the region of the possible phase coexistence, see Fig.~\ref{fig1},
 occupies significant part of the phase diagrams.

{\em First order phase transition at $d=2$}--
The extrapolation of RG equations to $%
\epsilon =1$ shows that for a weak Coulomb interaction,
$g_{1}^{d=2}\gg 1$, 
RG flow  has a stable fixed point, {\em i.e.} 
 the SIT is of the second order. Indeed, for this case
$g_1(l)=2l/3+const$, $g_2\to \epsilon/6$.  Then, $b$ tends to its fixed
point value as $b\approx 4\epsilon/5+{\cal O}(1/g_1)$ and the effect of the Coulomb interaction
vanishes at $l\to \infty$. Therefore, the effect
of the Coulomb interaction may lead only to the logarithmic corrections to
the usual power laws of $d=3$ classical $XY$ model.

Situation changes in the opposite limit when  the
mutual capacitance significantly exceeds the intergrain distance
i.e. $g_2 \gg g_1$. We can set in Eqs.~(\ref{a25})
$g_1=0$, then  $g_2$ rapidly reaches its fixed point $5/6\epsilon$ and
for $\epsilon=1$  evolution of $b(l)$ is governed by function (\ref{F}):
\begin{equation}
b(l)={\cal F} \left(l+l_* \right);\ b_*={\cal F} \left(l_* \right),
\label{b2D}
\end{equation}
where $b_* \sim bg_2$. Accordingly,
$b$ always changes sign and the SIT is of the first order at 
$T<T_t\simeq E_0e^{-|l_*|}$, see Fig.~\ref{fig1}b. If the Coulomb
interaction is moderate, $g_2\simeq g_1$, both kind of phase
transitions are possible, see Fig.~\ref{fig1}c. 

{\em Physical interpretation} -- The stability of both insulating
and superconducting states at $T<T_t$ appears due to the competition of the
effects of the long range
interaction of excitations against their Bose statistics. 
When the former is strong enough there exists a
state formed by boson/antiboson dipoles.
This stable state competes with the formation of the uniform Bose-condensate.

{\em In conclusion} we have derived an effective field theory describing the
SIT transition in granular superconductors and JJA. 
The RG analysis of this model demonstrated that the SIT is inevitably of first order
at low temperatures for all 3-dimensional and realistic 2-dimensional JJA. This may have very important experimental
consequences. 
In particular, one can see hysteresis
when changing {\em e.g.} magnetic field. The insulating and
superconducting state can coexist
and phase separation in space, especially in the presence of disorder,
is possible. 
All these interesting phenomena deserve a separate study.

Support by  US DOE contract No. DE-
AC02-06CH11357 (I.L.A. and B.L.A.) and Transregio 12 of DFG (
S.V.S. and K.B.E.) is gratefully acknowledged.

\end{document}